\documentclass[%
 reprint,longbibliography,
 superscriptaddress,
 amsmath,amssymb,
 aps,
floatfix,
]{revtex4-1}

\usepackage{graphicx}
\usepackage{dcolumn}
\usepackage{bm}
\usepackage[separate-uncertainty = true,multi-part-units=single]{siunitx}
\usepackage{tabularx}
\usepackage[section]{placeins}
\usepackage[dvipsnames]{xcolor}

\usepackage{xcolor}
\usepackage[xcolor=dvipdf,ulem=normalem,markup=underlined]{changes}
\definechangesauthor[color=red]{PVS}
\definechangesauthor[color=blue]{MY}
\definechangesauthor[color=purple]{AHM}

\newcommand{\mcite}[1]{\mbox{\cite{#1}}}

\newcommand{\nanometer}[1]{\SI{#1}{\nano\metre}}
\newcommand{\micron}[1]{\SI{#1}{\micro\metre}}
\newcommand{\ev}[1]{\SI{#1}{\electronvolt}}
\newcommand{\mev}[1]{\SI{#1}{\milli\electronvolt}}
\newcommand{\textsub}[1]{\textsubscript{#1}}
\newcommand{\commenta}[1]{}
\newcommand{\commentb}[1]{}

\newcommand{\lSAW}{{\lambda_{\mathrm{SAW}}}}		

\begin{document}

\preprint{APS/123-QED}

\title{Sidewall quantum wires on GaAs(001)  substrates}

\author{Paul L. J. Helgers} 
\email{helgers@pdi-berlin.de}
\affiliation{Paul-Drude-Institut f{\"u}r Festk{\"o}rperelektronik, Leibniz-Institut im Forschungsverbund Berlin e. V., Hausvogteiplatz 5-7, 10117 Berlin, Germany }
\affiliation{NTT Basic Research Laboratories, NTT Corporation, 3-1 Morinosato-Wakamiya, Atsugi, Kanagawa 243-0198, Japan}

\author{Haruki Sanada}
\author{Yoji Kunihashi}
\affiliation{NTT Basic Research Laboratories, NTT Corporation, 3-1 Morinosato-Wakamiya, Atsugi, Kanagawa 243-0198, Japan}

\author{Klaus Biermann}
\author{Paulo V. Santos}
\affiliation{Paul-Drude-Institut f{\"u}r Festk{\"o}rperelektronik, Leibniz-Institut im Forschungsverbund Berlin e. V., Hausvogteiplatz 5-7, 10117 Berlin, Germany }

\date{\today}

\begin{abstract}
We study the structural,  optical, and transport properties of sidewall quantum wires on GaAs(001) substrates. The QWRs are grown by molecular beam epitaxy (MBE) on  GaAs(001) substrates pre-patterned with shallow ridges. They form as a consequence of material accumulation on the sidewalls of the ridges during the  overgrowth of a quantum well (QW) on the patterned surface. The QWRs are approximately 200~nm-wide and have emission energies red-shifted by \mev{27} with respect to the surrounding QW. Spatially resolved spectroscopic photoluminencence studies indicate that the QW thickness reduces around the QWRs, thus creating a  4 meV energy barrier for the transfer of carriers from the QW to the QWR. We  show that the QWRs act as efficient channels for the transport of optically excited electrons and holes  over  tens of $\mu$m by a high-frequency surface acoustic wave (SAW). These results demonstrate the feasibility of efficient ambipolar transport in  QWRs with sub-micrometer dimensions, photolithographically defined on GaAs substrates. 

\end{abstract}

\pacs{Valid PACS appear here}
\maketitle

\section{Introduction}
Planar quantum wires (QWRs) are important components for the realization of interconnects for integrated optical-electronic circuits. In its simplest form, the guiding  action by the lateral confinement of carriers enables information exchange between two remote locations via a well-defined channel. The mesoscopic confinement induced by small lateral dimensions can be explored for additional functionalities. Here, interesting examples are the formation of one-dimensional quantum channels for electronic transport with reduced scattering~\cite{Fasol_prl70_3643_93} as well as the enhanced spin lifetimes due to mesoscopic spin confinement~\cite{Kiselev_PRB61_13115_00,Holleitner_PRL97_036805_06}. 

Different approaches have been reported for the fabrication of semiconductor QWRs with nanometer dimensions.  Electrostatically defined QWRs can be created via the deposition of gates on a semiconductor nanostructure. The application of a voltage to the gate  creates channels near the surface for one type of carriers (i.e., electrons or holes). Ambipolar transport channels for the guidance of both electrons and holes by moving acoustic fields can be created by combining electrostatic gates with piezoelectricity \cite{PVS290}. More interesting for electro-optical applications are planar QWRs defined by a lateral structural modulation since these have optical resonance energies distinct from the ones of the surrounding matrix. In its simplest form, the structural modulation can be introduced by etching a QW sample to define the QWR. Alternatively, a QWR can be formed by the epitaxial growth of a QW structure on a surface with a pre-defined structural modulation. Here, one example is given by QWRs produced by the epitaxial growth on the cleaved edge of a nanostructure containing a QW ~\cite{Pfeiffer_JCG127_849_93}. A second example is provided by the epitaxial growth on a surface exposing different crystallographic orientations: here, one takes advantage of the dependence of the growth kinetics on the orientation of the surface in order to create the lateral structural modulation. Surface structuring in this case is usually achieved by pre-etching the substrate  surface to expose different crystallographic facets. Well-known examples are QWRs grown by metal-organic epitaxy on V-shaped grooves defined on a GaAs(001) surface \cite{Bhat_JCG93_850_88,Kapon89a,Koshiba_APL64_363_94} as well as quantum wires and dots fabricated by molecular beam epitaxy (MBE) on pre-patterned GaAs surfaces with different orientations~\cite{Horikosh_Book_27_18,Noetzel_APL68_1132_96,Noetzel_JAP40_4108_96,Noetzel98c}. An advantage of these growth-defined QWRs over etched ones resides on the fact that they do not contain free surfaces, which may be deleterious for electronic excitations. In addition, and in contrast to their electrostatic counterparts, these QWRs are fully surrounded by epitaxial materials and can therefore be easily embedded in more complex epitaxial structures such as optical microcavities.

\begin{figure*}[htbp]
\includegraphics[width=.85 \textwidth, angle=0, clip]{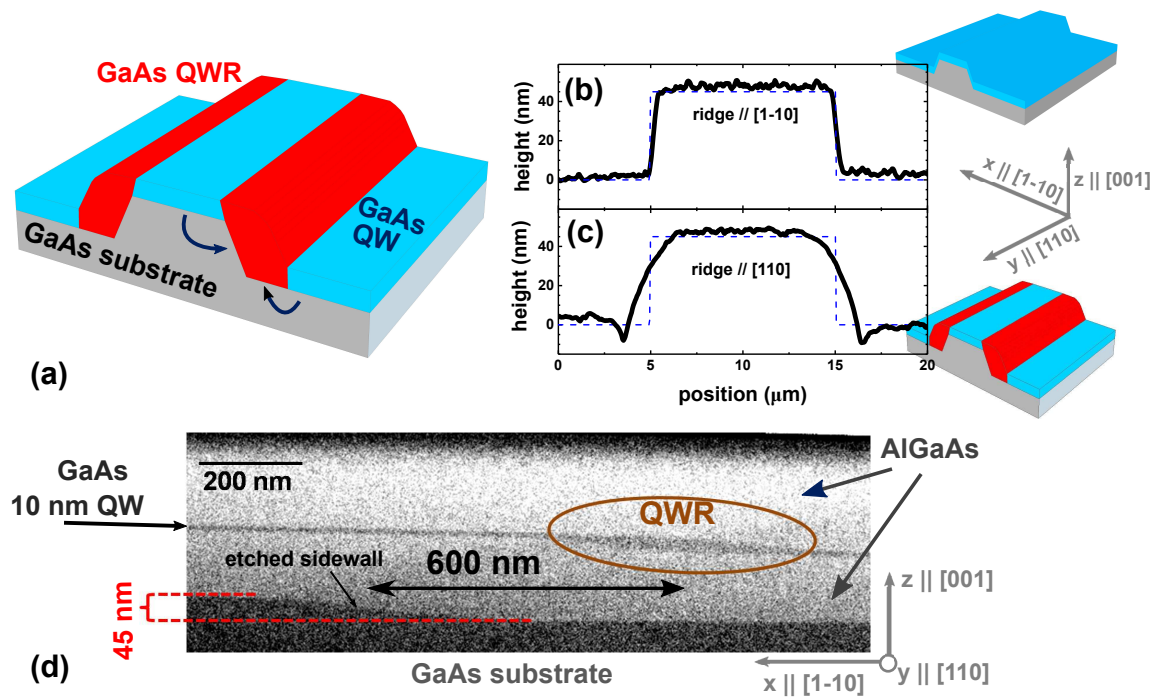}
\caption{Molecular beam epitaxy (MBE) of sidewall quantum wires (QWR) on GaAs(001). (a) Shallow ridges (height of approx. \nanometer{45}) were patterned on a GaAs(001) substrate and subsequently overgrown with a quantum well (QW, blue areas). The arrows show the preferential diffusion of Ga adatoms towards the ridge sidewall, which results in the QWR formation (red regions). For clarity the AlGaAs QW barriers have been omitted. (b)-(c) Atomic force microscopy (AFM) cross-sectional profiles of overgrown  ridges oriented along the (b) $x || [1-10]$ and (c) $y || [110]$ directions of the GaAs(001) surface. The blue dashed lines indicate the contour of pre-patterned ridges defined by optical lithography. The pronounced lateral growth results in  QWR formation only along the $y$-oriented sidewalls.  (d) Cross-sectional scanning transmission electron micrography (STEM) image of an overgrown ridge showing the QW (GaAs, dark layer) sandwiched between two AlGaAs barriers (bright layers). The brown circle highlights the region of increased QW thickness defining the sidewall QWR. 
}
\label{fig:Fig1}
\end{figure*}

One important application of planar QWRs is as guides and interconnects for photo-excited electron-hole pairs between opto-electronic structures on a surface. The charge carriers can be stored and transported along the plane of a QW by the traveling piezoelectric potential of a surface acoustic wave (SAW) \cite{Rocke97a,PVS090}. After transport, the carriers can be forced to recombine and emit photons, thus providing a photonic interface to the electronic system. Similar ambipolar transport experiments were demonstrated in $\mu$m-sized electrostatically defined wires \cite{PVS229} as well as sub-$\mu$m sidewall QWRs fabricated by epitaxially overgrowing shallow ridges on  GaAs(113)A substrates \cite{PVS124,PVS129}.  Long acoustic driven transport distances for charge carriers have been achieved in overgrown QWRs (up to $50~\mu$m) \cite{PVS192}. Their optical and transport properties were found, however,  to be very sensitive to  potential fluctuations along the transport path. These fluctuations create trapping centers, which can capture charge carriers during the acoustic driven ambipolar transport and induce their recombination, thus resulting in a reduction of the acoustic transport efficiency \cite{PVS135}. 

The one-directional motion of carriers in QWRs also provides a pathway to reduce spin dephasing due to the Dyakonov-Perel mechanism ~\cite{Kiselev_PRB61_13115_00,Holleitner_PRL97_036805_06}. Alsina et al. \cite{PVS192} investigated spin transport in  sidewall QWRs grown on GaAs(113)A substrates with a width of \nanometer{50}. Whereas the narrow width should lead to a long spin relaxation time and,  correspondingly, long spin transport lengths, the observed acoustic spin transport length was limited to distances of only approx. \micron{2}. These transport distances are much shorter than the  ones reported for acoustic transport in electrostatic wires \cite{PVS265}. It was argued that the acoustic spin transport length is limited by Elliot-Yafet (EY) scattering due to the large density of scattering centers along the QWR axis \cite{PVS192}. These findings call for fabrication processes yielding narrow QWRs with small potential fluctuations.

The formation of sidewall QWRs via MBE overgrowth on patterned GaAs(001) substrates was reported by Lee et al.~\cite{Lee_ITN6_70_07}. From a fabrication point of view, the MBE growth process on the GaAs(001) surface is better understood and controlled than on high-index surfaces such as GaAs(113)A. Previous studies of sidewall QWRs on GaAs(001), however, only addressed structural properties. Neither their optical nor transport properties have so far been investigated.

In this contribution, we provide systematic investigation of the structural, optical, and acoustic transport properties of sidewall QWRs fabricated on GaAs(001) substrates. The sample fabrication process, which includes substrate patterning and MBE growth, was based on the previous work in Ref.~\onlinecite{Lee_ITN6_70_07} and is summarized in Section \ref{sec:fabrication}. Sections~\ref{Excitation of surface acoustic waves} and \ref{Optical spectroscopy techniques} then describe the procedures for the fabrication of interdigital transducers for SAW generation and the spectroscopic techniques employed in the studies. The experimental results are presented in Sec.~\ref{sec:results}. Here, we start by studying the structural properties  of the QWRs by combining atomic force microscopy (AFM),  scanning electron microscopy (SEM), and scanning transmission electron microscopy (STEM) (Sec.~\ref{sec:structural}).   We then carried out spectroscopic investigations of the QWR optical properties using spatially resolved photoluminescence (PL) (Sec.~\ref{sec:optical}-D). Finally, Sec.~\ref{sec:transport} provides evidence for the acoustic charge transport in the QWRs over distances approaching $100~\mu$m. Section \ref{sec:conclusion} summarizes the main conclusions drawn in this work.

\section{\label{sec:Experimental_Details}Experimental details}

\subsection{\label{sec:fabrication} Sample fabrication}

The fabrication of the sidewall QWRs on GaAs(001) substrates followed a procedure similar to the one reported for QWRs reported in Refs.~\onlinecite{Noetzel_APL68_1132_96} (GaAs(113)A) and \onlinecite{Lee_ITN6_70_07} (GaAs(001)). In the first step, shallow ridges  were  patterned on a GaAs(001) substrate by photolithography and wet chemical etching using a solution of H\textsubscript{2}SO\textsubscript{4}:H\textsubscript{2}O\textsubscript{2}:H\textsubscript{2}O with a volume ratio of (8:1:100). We fabricated \micron{10}-wide ridges with a height of \nanometer{45} and a length of several tens of \SI{}{\micro\metre} oriented along the $y||[110]$ and $x||[1-10]$ main axes of the GaAs(001) surface.
The patterned substrate was subsequently cleaned using a H\textsubscript{2}SO\textsubscript{4}:H$_2$O (96:4) solution and introduced in a UHV chamber connected to the MBE growth apparatus for surface cleaning by exposure to  atomic hydrogen. In this procedure, the substrate was exposed to partially cracked hydrogen from a hot filament source at a background pressure of \SI{5e-5}{\milli\bar{}} for 30 minutes at a temperature of \SI{450}{\celsius}. The sample was then transferred in vacuum to the MBE growth chamber. Figure~\ref{fig:Fig1}(a)  schematically shows the formation of QWRs (red areas) after the overgrowth of a layer structure, consisting of a \nanometer{10} QW (blue) sandwiched between a lower and upper Al\textsub{0.15}Ga\textsub{0.85}As barrier of respectively \nanometer{130} and \nanometer{200} thickness. For comparison, a sample containing QWRs on a GaAs(113)A substrate has been fabricated following a similar procedure, using an H\textsubscript{2}SO\textsubscript{4}:H\textsubscript{2}O\textsubscript{2}:H\textsubscript{2}O etching solution with a volume ratio of (1:8:100). For both types of samples, all layers were grown while the substrate was kept at a temperature of \SI{600}{\celsius}. The growth rate of the AlGaAs and GaAs layers was respectively \SI{0.16}{\nano\metre\per\second} and \SI{0.14}{\nano\metre\per\second}. Finally, the samples were capped with a \nanometer{2} layer of GaAs in order to protect the sample against oxidation.

The formation of QWRs on the ridge sidewalls relies on the combination of two factors \cite{Lee_ITN6_70_07,Mirin_JVSTA10_697_92,Horikoshi_Book_1993,LaBella_JVSTA18_1526_00}: (i) the higher adatom diffusion rate along the  $x||[1-10]$ surface direction of the (2$\times$4)  reconstructed GaAs(001) surface as compared to the $y||[110]$ direction and (ii) the higher adatom  incorporation rate at step edges oriented along $y$ as compared to the plane (001) surface.
The latter results in an enhanced MBE growth rate on the exposed sidewalls of the ridges as compared to the growth rate on the (001) surface. When a QW is deposited on the patterned surface, the anisotropic growth rate induces a local increase in the QW thickness on the sidewalls of ridges aligned along the $y$ direction, as indicated by the red regions in Fig.~\ref{fig:Fig1}(a) ~\cite{Lee_ITN6_70_07}.
The structural properties of the overgrown ridges were investigated by combining AFM, SEM and STEM.

\subsection{Excitation of surface acoustic waves}
\label{Excitation of surface acoustic waves}

\begin{figure}[tbhp]
\includegraphics[width=\columnwidth, angle=0, clip]{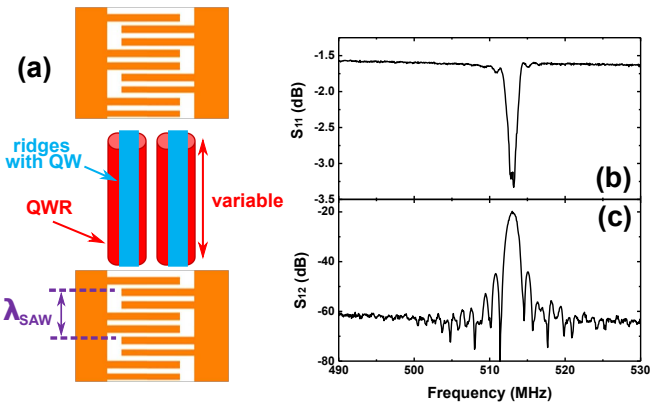}
\caption{(a) Schematic diagram of the SAW delay line embedding sidewall QWRs aligned along the $y||[110]$ direction of the GaAs(001) substrate. The delay line consists of two interdigital transducers (IDTs) with an aperture of \micron{120} and acoustic wavelength $\lSAW=$\micron{5.6}. (b) Radio-frequency (RF) reflection ($s_{11}$) and (c) RF transmission ($s_{21}$)  parameters for the delay line measured at room temperature. The resonance at \SI{513}{\mega\hertz} is associated with the excitation of the Rayleigh SAW mode of the sample structure.
}
\label{fig:Fig2}
\end{figure}

Figure \ref{fig:Fig2}(a) displays the layout of the  delay line used for the excitation of Rayleigh SAWs \cite{White_APL7_314_65} along the QWR axis. The delay line was fabricated by optical lithography on the surface of the overgrown sample: it  consists of two split-finger interdigital transducers (IDTs) designed for an acoustic wavelength $\lSAW= \micron{5.6}$ with aperture and length of \micron{120} and $350\times\lSAW$, respectively.
 Figures~\ref{fig:Fig2}(b) and \ref{fig:Fig2}(c) show the frequency dependence of the radio-frequency (rf) reflection ($s_{11}$)  and transmission ($s_{21}$)  parameters at room temperature, respectively. The resonance at  \SI{513}{\mega\hertz} corresponds to the excitation of the Rayleigh mode for the structure. The amplitude of the $s_{11}$ dip indicates that the IDTs convert 20\% of the input rf-power into two SAW modes propagating in opposite directions. The $r_\mathrm{IDT}=10\%$ transduction per SAW beam is compatible with the maximum power transmission amplitude of the $s_{12}$ spectrum of -20~dB. 
 
\subsection{Optical spectroscopy techniques}
\label{Optical spectroscopy techniques}
The spectroscopic photoluminescence (PL) studies were carried out using an optical microscope coupled to a cryostat (temperature of approx. \SI{10}{\kelvin}) with optical access and rf-wiring for the excitation of SAWs.  The PL was excited by either a continuous wave (cw) or pulsed tunable Ti:Sapphire laser, or by a pulsed diode laser. The PL around the excitation spot was spectrally analysed and detected with a spatial resolution of \micron{0.5} per pixel by a cooled charged-couple device (CCD) camera. Time-resolved measurements were performed using either a streak-camera or an avalanche Si-photodiode as a detector.
 
The acoustic transport studies  were carried out by exciting one of the IDTs of the delay line. The electron-hole pairs were excited by a focused laser spot on one position of the sample (cf. Fig.~\ref{fig:Fig2}(a)) and their spatial distribution along the SAW propagation direction detected by recording spatially resolved PL profiles. The amplitude of the SAW will be stated in terms of the nominal rf-power applied to the rf-input of the cryostat. Note that the electro-acoustic conversion efficiency in the cryostat will be lower than the one  determined in Fig.~\ref{fig:Fig2}, which does not take into account the effects of the cryostat rf-connections.

\section{Results and discussion}
\label{sec:results}

\subsection{ Structural properties}
\label{sec:structural}

Since MBE  is a non-conformal growth technique, the shape of the etched ridges can be recovered by probing the sample surface after overgrowth using AFM. 
The solid lines in Figs.~\ref{fig:Fig1}(b) and \ref{fig:Fig1}(c) compare cross-sectional AFM profiles of overgrown ridges oriented along the $x$ and $y$ directions, respectively. The blue dashed lines indicate the nominal profiles of the ridge etched on the surface prior to the MBE overgrowth. The cross-sectional profile of overgrown ridges oriented along the $x$-direction closely follows the one pre-patterned on the surface, thus indicating a conformal coverage during the MBE overgrowth.   Ridges oriented along $y$ exhibit, in contrast, broadened sidewalls with a convex shape. This behaviour is attributed to the  higher growth rates at the edge of these ridges, which results in material accumulation along their sidewalls. This material accummulation locally increases the thickness of a QW overgrown on the ridge edges, thus forming the region with lower quantum confinement energy, corresponding to the QWR (cf. Fig.~\ref{fig:Fig1}(a)).

The material transfer leading to QWR formation also reduces the thickness of the QW regions adjacent to the sidewalls. As will be discussed in detail in Sec.~\ref{sec:optical}, this local thickness decrease creates a ``barrier-QW'' with higher carrier confinement energies than in the QW on both sides of the QWR. These ``barrier QWs'' act as potential barriers for the transfer of carriers from the QW to the QWR.

The sidewall QWRs only form on overgrown ridges oriented along $y$. These can be directly imaged in cross-sectional STEM images, as illustrated in Fig.~\ref{fig:Fig1}(d). Here, the dark and bright areas correspond to GaAs and Al\textsub{0.15}Ga\textsub{0.85}As regions, respectively. From the micrographs, we estimate the thickness and width of the sidewall QWR to be \SI{25\pm5}{\nano\metre} and  \nanometer{200\pm5}, respectively. The QWR is thus approx. 2.5 times thicker than the deposited \nanometer{10} QW. Interestingly, the QWR does not form directly on the location of the ridge sidewall on the substrate, but is shifted by approximately \nanometer{600}. From the ratio between the \nanometer{600} lateral shift and the \nanometer{130} thickness of the lower Al\textsub{0.15}Ga\textsub{0.85}As barrier, we estimate that for this particular geometry the lateral growth rate of the Al\textsub{0.15}Ga\textsub{0.85} layer is approximately 5 times larger than the vertical one. 

\begin{figure*}[htbp]
\includegraphics[width=1\textwidth, angle=0, clip]{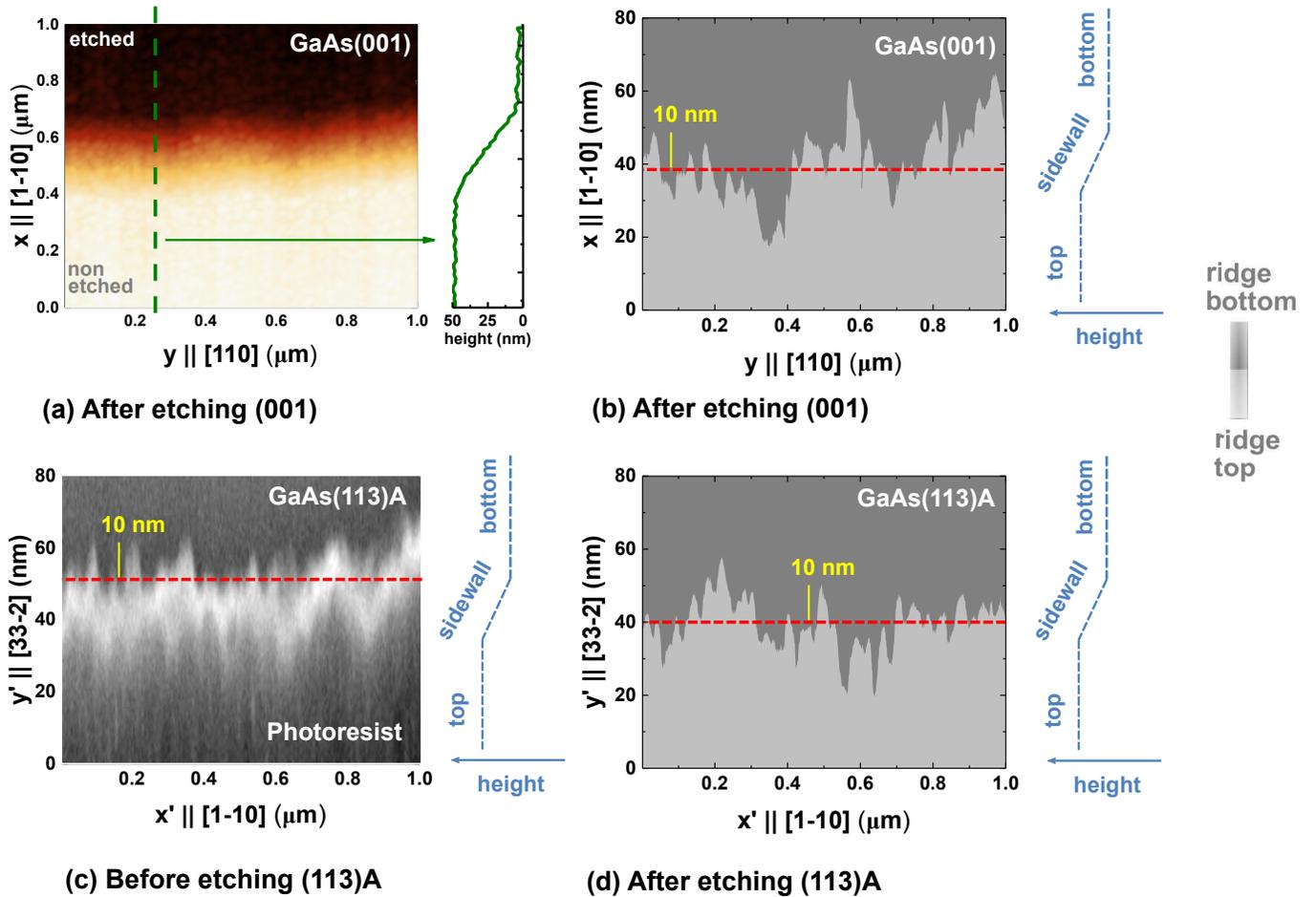}
\caption{Line-edge-roughness of ridges at GaAs. (a) AFM scan of a \micron{1} $\times$ \micron{1} sized area around a ridge sidewall after etching and before MBE-overgrowth. The right panel shows a profile extracted along the green dashed vertical line. (b) Post-processed image of the same scan, zoomed into a \micron{1} $\times$ \nanometer{80} sized area. The ridge-bottom is dark gray coloured and its top is bright gray coloured. The red dashed line depicts the average position of the step edge. The value of the line-edge-roughness is \nanometer{9}. (c) Scanning electron microscopy image of the photoresist defining the ridges after resist exposure to UV light of \nanometer{250} (but before etching). The upper part of the image corresponds to a GaAs(113)A substrate and the lower part of the image corresponds to the photoresist. The rms LER value of  \nanometer{10} is  comparable to the deviations observed in the etched ridge shown in figure (d). (d) Post-processed image of an AFM scan of a \micron{1} $\times$ \nanometer{80} sized area around a ridge sidewall on a GaAs(113)A substrate after etching and before MBE overgrowth. The LER of etched ridges on GaAs(113)A (approx. \nanometer{7} for the shown measurement) has a similar amplitude as LER on GaAs(001).
}
\label{fig:Fig3}
\end{figure*}

\begin{figure*}[htbp]
\includegraphics[width=\textwidth, angle=0, clip]{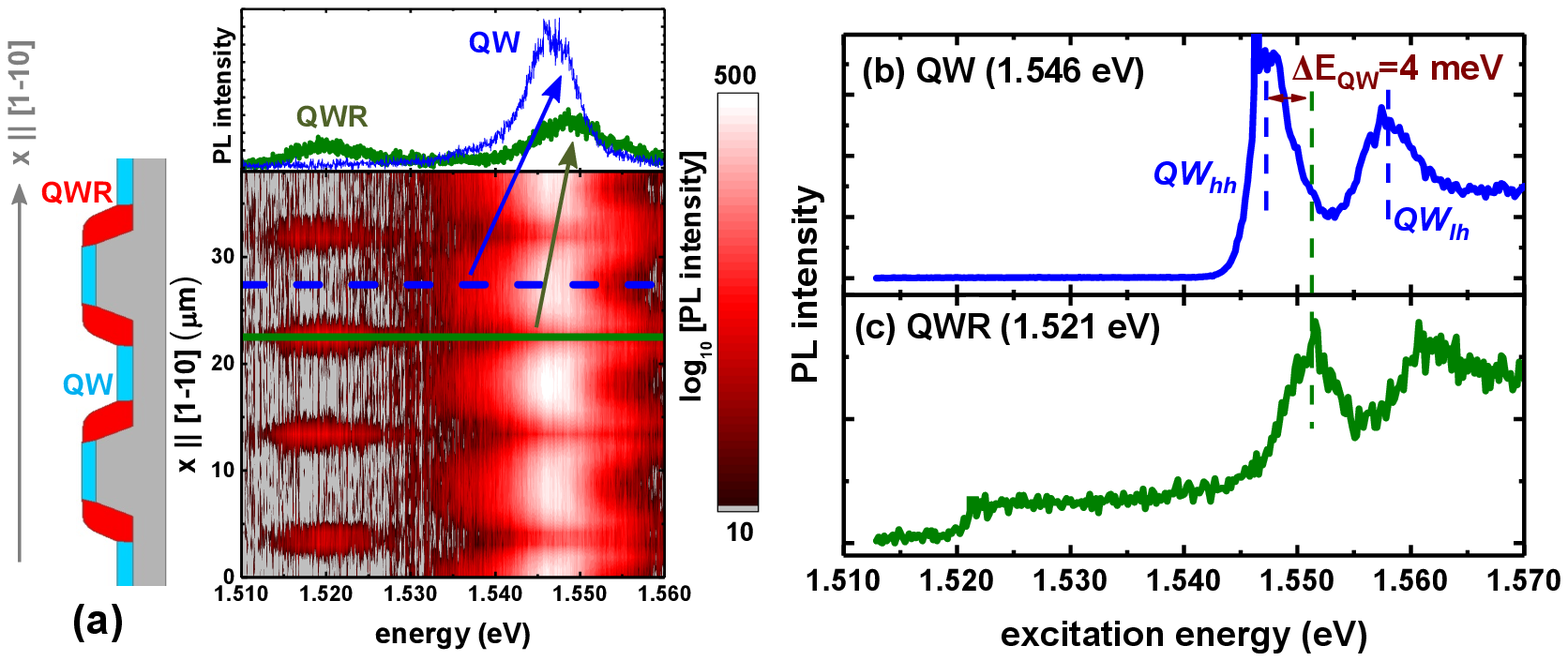}
\caption{(a) Photoluminescence (PL) map recorded while scanning the excitation laser spot in the direction perpendicular to the \micron{10}-wide  ridges with sidewall QWRs (cf. left panel).  The map shows the colour-coded  PL intensity (in a logarithmic scale) as a function of position (vertical axis) and emission energy (horizontal axis). Emission is observed from two distinct resonances. 
The low energy corresponds to the QWR emission and the high energy corresponds to the QW. 
The upper panel shows spectra along the cross-sections of the PL map indicated by the horizontal lines, where the QW spectrum depicts the spectrum extracted at the blue dashed line position. Photoluminescence excitation (PLE) spectra detected at the electron-heavy hole resonance are shown in figures (b)  QW (QW$_\mathrm{hh}$) and (c) QWR (QWR$_\mathrm{hh}$). In both cases, the  excitation spot was placed on a sidewall QWR. 
}
\label{fig:Fig4}
\end{figure*}

The electro-optical properties and, in particular, the carrier transport properties of the sidewall QWRs also depend on the uniformity of the QWR dimensions along both the growth direction and the QWR axis direction. The latter can be quantified  by evaluating the lateral roughness of the ridge edges (line-edge-roughness, LER) from   AFM micrographs. For that purpose, we have recorded AFM profiles across patterned ridges aligned along $y$ (typically one profile per nanometer). The constructed image of a ridge sidewall  wet-chemically etched on the GaAs(001) substrate (i.e., prior to MBE overgrowth) is shown in Fig. \ref{fig:Fig3}(a). Here, the dark area corresponds to the ridge bottom (etched) and the bright area to the ridge top (non-etched). The right panel displays a typical profile, extracted along the vertical dashed green line in the AFM image. Each of these profiles was fitted with an error function to evaluate the average position of the ridge edge. LER is defined as the standard deviation of the edge position. Figure~\ref{fig:Fig3}(b) shows the same scan as Fig. \ref{fig:Fig3}(a), now zoomed into an area of \micron{1} x \nanometer{80}. The image was post-processed to show only the heights of the ridge top and bottom. The edge position, which is depicted by the line dividing the top and bottom areas, fluctuates with peak amplitudes of as much as \nanometer{20} and shows a LER value of \nanometer{9}. Furthermore, we observe that the sidewall of the etched ridge is inclined by an angle of roughly \SI{10}{\degree} with respect to the surface plane.

Interestingly, the LER of ridges etched on GaAs(001) QWs was found to be approximately the same as in control samples deposited on GaAs(113)A substrates (cf. Fig. \ref{fig:Fig3}(d)): in both cases, we measured LER in the range from \nanometer{4} to \nanometer{9}. Also,  the LER could not be reduced by changing the composition of the etching solution (i.e., the type of acid agent as well as  the degree of dilution of the etching solvent). We found, in contrast, that the  LER can be traced back to the photolithography process used to define the ridges. In fact, SEM images of the photoresist edges at GaAs(113)A (cf. Fig.~\ref{fig:Fig3}(c)) display LER values very similar to the ones measured after overgrowth (cf. Fig.~\ref{fig:Fig3}(d)). Several factors leading to the LER of photoresist patterns are   discussed in detail in Ref.~\onlinecite{MACK_SPIE_2010}. 

Attempts were also carried out to improve the LER of the sidewalls using patterns defined by electron-beam lithography. Ridges defined on a GaAs(113)A substrate by e-beam lithography using a step size of \nanometer{20} and an Allresist AR-P 6200 resist have LERs similar to the ones obtained by optical lithography. By reducing the stepsize of the e-beam lithography to \nanometer{2} and using a PMMA resist, the LER could be reduced to approximately \nanometer{3}, thus indicating a way to improve the ridge quality. 

In the present studies, however, all QWRs were fabricated on photolithographically defined ridges. Note that since the LER is only a few percent of the width of the sidewall QWR determined by STEM (see Fig.~\ref{fig:Fig1}(d)), we expect that it will only play a minor role in the optical, electrical and transport properties of the sidewall QWRs. 

\subsection{Optical properties of the QWRs}
\label{sec:optical}

The formation of QWRs at the ridge sidewall was verified by spatially resolved PL spectroscopy.
Figure~\ref{fig:Fig4}(a) displays a PL map recorded at \SI{10}{\kelvin} while scanning a laser spot across ridges with sidewall QWRs (cf. left panel).  We used for excitation a pulsed diode laser emitting at a wavelength of \nanometer{635}. The latter lies above the fundamental optical transitions of the QW and, thus, optically excites carriers  both in the QW and the QWR.  The scan length covers two $10~\mu$m-wide overgrown ridges separated from each other by $10~\mu$m.  The map was generated by  spectrally analyzing, for each spot position,  the PL emitted within a \micron{1.5} x \micron{2} area around it. The procedure yields the PL map with the color-coded intensity in a logarithmic scale as a function of position (vertical axis) and energy (horizontal axis).  The upper panel displays spectral cross-sections of the map recorded along the blue dashed (QW at ridge top) and green solid (sidewall QWR) horizontal lines indicated in the map.

The PL map of Fig.~\ref{fig:Fig4}(a) shows two main resonances. Spectra recorded on the top and bottom of the ridges are essentially equal and characterized by a single PL line  centered at  \ev{1.547} with a linewidth (defined as  the full-width-at-half-maximum (FWHM)) of \mev{7}. This line is  attributed to the electron-heavy hole (QW$_\mathrm{hh}$) resonance of the \nanometer{10} QW. Spectra recorded on the sidewall display, in addition, a second resonance at \ev{1.520} with a  linewidth of \mev{10.5}, which is assigned to the electron-heavy hole transition of the QWR (QWR$_\mathrm{hh}$). As was discussed in Section \ref{sec:structural}, the width of the QWR is much larger (approximately \nanometer{200}) than its thickness (approximately \nanometer{25}). The red-shift of the QWR emission with respect to the QW is, therefore, mainly caused by its larger thickness with respect to the QW. Furthermore, due to its increased thickness, one expects a decrease of the QWR linewidth with respect to the QW. The opposite trend observed in the experiments implies that potential fluctuations within the QWR have a significant impact, possibly related to the aforementioned LER. The fact that the QWR$_\mathrm{hh}$ resonance is only seen at the sidewall positions (see left panel) also confirms the structural findings of Sec.~\ref{sec:structural} that the QWRs only form at the sidewalls. 

A closer analysis of the PL spectra recorded at the sidewalls reveals that the QW line reduces in intensity and slightly blue-shifts (by approx. \mev{3}) with respect to spectra taken at a position away from the ridges (not shown here). The blue-shifted region is attributed to the ``barrier QWs'' with reduced thickness formed on both sides of the sidewall QWR.  Electron-hole pairs photo-excited within the barrier QW can diffuse to and recombine in the neighboring regions of lower energy. The blue-shifted PL emission  at the ridge sidewall in Fig.~\ref{fig:Fig4}(a) is symmetric with respect to the axis of the QWR, thus proving that the barrier QW forms an  energy barrier for the transfer of carriers from the QW to the QWR on both sides of the QWR. Considering that the dimension of the sidewalls is much smaller than the optical resolution, the PL line in Fig.~\ref{fig:Fig4}(a)  includes contributions from both the barrier-QW and the QW. The latter may be responsible for the increased linewidth of the QW PL to \mev{11} on the sidewall as compared to \mev{7} measured away from the sidewall.

\subsection{Comparison with other sidewall QWRs}
\label{Sec:Comparison with other planar QWRs}

Table \ref{table:linewidths} compares the optical properties of sidewall QWRs grown on GaAs(001) substrates with QWRs grown on GaAs(113)A substrates. 
The table was constructed using average values of the emission energies and linewidths of QWRs from multiple samples simultaneously grown the same 2-inch wafer, per substrate type.
The fourth column displays the emission energies and linewidths of QWRs fabricated on electron-beam lithographic defined ridges on a GaAs(113)A substrate overgrown with a slightly thinner QW (\nanometer{8} instead of \nanometer{10}). 
Interestingly, all sidewall QWRs have similar spectral linewidths of $10\pm1$~nm.  
The energetic separation $\Delta E_\mathrm{QWR}$ between the QW and the QWR emission energies, in contrast, is significantly larger for QWRs on GaAs(001) than on GaAs(113)A and leads to a larger carrier confinement for GaAs(001). This  larger red-shift, which is probably related to the different sidewall facets exposed by the etching process,  is advantageous for acoustic carrier transport since it reduces the carrier escape probability to the  surrounding QW. Finally, the last column of the table list the corresponding properties for V-groove QWRs deposited by Koshiba et al. on GaAs(111) substrates (from Ref.~\onlinecite{Koshiba_APL64_363_94}). These QWRs have a linewidth comparable to the ones for the sidewall QWRs but substantially larger energetic separations $\Delta E_\mathrm{QWR}$.

\begin{table}
\begin{tabular}{|l|c|c|c|c|}
\hline
&&&&\\
       & (001) & (113)A  	& (113)A  & (111)\\         
       &  		&   			& e-beam & V-groove\\         
\hline
QW width (nm)      &     10      &     10      &  8  & 7  \\
E\textsub{QWR} (eV)      &     1.520      &     1.526      &  1.546  & 1.575  \\
$\Delta E_\mathrm{QWR}$ (meV) &    27       &       15     & 13  & 45 \\
linewidth\footnote{defined as the FWHM} (meV)    &      10     &      11    & 9 & 11\\
\hline
\end{tabular}
\caption{Comparison of the optical properties of sidewall QWRs on (001) and (113) GaAs substrates. The last column list the corresponding properties for V-groove QWRs on (111 ) (from Ref.~\onlinecite{Koshiba_APL64_363_94}). $\Delta E_\mathrm{QWR}$ denotes the red-shift of the QWR PL line with respect to the QW line.}
\label{table:linewidths}
\end{table}
  
 
 \subsection{Carrier dynamics in GaAs(001) QWRs}
 
In order to investigate carrier transfer between the QW and the QWR, we measured photoluminescence excitation spectra (PLE) from the sidewalls  using a tunable cw Ti:Sapphire laser impinging on the sidewall region at an angle of incidence of \SI{45}{\degree}. The laser spot size of approximately \micron{7} in diameter is much larger than the sidewall and can, thus, generate carriers both in the QWR and in the QWs around it.
Figures \ref{fig:Fig4}(b) and \ref{fig:Fig4}(c) compare PLE spectra by detecting the emission of the  QW$_{hh}$  (detection energy \ev{1.546}) and of the QWR$_\mathrm{hh}$ (\ev{1.521}) resonances, respectively. In the former case, the two peaks in the emission intensity are attributed to the electron-heavy hole (QW$_\mathrm{hh}$) and electron light-hole (QW$_\mathrm{lh}$) transitions of the QW  at  \ev{1.548} and \ev{1.558}, respectively. This assignment is consistent with calculations of the energy levels for a 10~nm-thick QW with Al$_{0.15}$Ga$_{0.85}$As barriers. 

The QWR$_\mathrm{hh}$ PLE spectrum in Fig.~\ref{fig:Fig4}(c)  also shows an onset at the QWR$_{hh}$ transition as well as two excitation maxima  blue-shifted by 4~meV and 5~meV with respect to the QW$_\mathrm{hh}$ and QW$_\mathrm{lh}$ transitions, respectively.

These blue-shifted lines are assigned to the electron-heavy hole and electron-light hole transitions in the barrier-QW. Carriers excited in the barrier-QW can then diffuse to the QWR and recombine in the QWR. The absence of a peak in the PLE spectrum of Fig.~\ref{fig:Fig4}(c) corresponding to the QW$_\mathrm{hh}$ transition as observed in Fig.~\ref{fig:Fig4}(b) implies that for excitation energies above approximately \ev{1.542}, the detected QWR PL arises from the recombination of electron-hole pairs photo-excited in the  barrier-QW region, which diffuse into the QWR. Carriers generated in the QW, in contrast, cannot reach the QWR due to the energy barrier imposed by the barrier-QW. For excitation energies below \ev{1.542} we expect that the QWR PL results from a combination of direct excitation of QWR states as well as of the low-energy flank of the barrier-QW.


\begin{figure}[htbp]
\includegraphics[width=1\columnwidth, angle=0, clip]{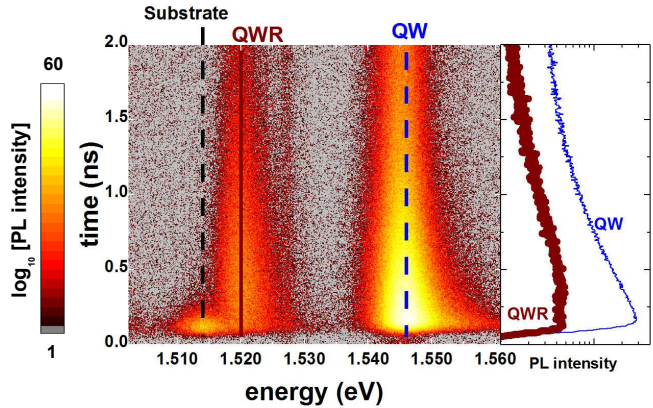}
\caption{Time-resolved photoluminescence traces recorded on a ridge sidewall as function of emission energy (horizontal axis) and time (vertical axis). The three observed resonances  are (from the low-energy to the high-energies) the electron-heavy hole transitions in the GaAs substrate (1.514 eV, dashed black line), sidewall QWR (1.520 eV, solid red line), and QW (1.546 eV, dashed blue line). The right panel shows the temporal trace of the QW (blue, thin) and QWR (red, thick). Both traces represent spectral integrations over the linewidth of the respective emission lines.}
\label{fig:Fig5}
\end{figure}


We analyze the carrier dynamics by performing time-resolved photoluminescence measurements using a streak camera while exciting the sample using \SI{1.5}{\pico\second} pulses from a Ti-Sapphire laser emitting at a wavelength of \nanometer{790}. Figure \ref{fig:Fig5} shows the PL intensity as function of energy (horizontal axis) and time (vertical axis) recorded by focusing a  \SI{3}{\micro\watt} laser beam onto a $7~\mu$m-wide laser spot on a sidewall under an angle of \SI{45}{\degree}. We observe emission from the sidewall QWR (\ev{1.520}),  QW (\ev{1.546}), as well as  from the GaAs substrate (\ev{1.514}). In contrast to the carriers in the QWR and QW, the PL signal from the substrate decays very fast  with an exponential decay time of \SI{70}{\pico\second}. The right panel shows temporal traces extracted from the image for the QW resonance (blue, thin line) and QWR (red, thick line)  by spectral integration of the PL over the linewidth of the corresponding resonances. In both cases, the formation of large-$k_{||}$ excitons from excited electron-hole pairs leads to the initial fast rise of the PL intensity and blue-shift of the emission line. These excitons then relax into a state with $k_{||}=0$ and subsequently recombine with the emission of a photon \cite{damen1990dynamics}. From a single-exponential fit we extract a QW carrier lifetime of approximately \SI{500}{\pico\second}.

The temporal emission trace of the QWR shows an initial fast increase followed by a plateau (or even a small increase) in the emission up to approximately \SI{400}{\pico\second}. The signal then decreases exponentially with a time decay constant of approximately \SI{870}{\pico\second}. The larger recombination lifetime of the QWR states with respect to the QW is attributed to the larger thickness of the QWR \mcite{Goebel_PRL51_1588_83}. We attribute the plateau to the transfer of carriers from the surrounding barrier-QW into the QWR. The initial rise time can be explained by the fast carrier transfer from the barrier-QWR as well as by direct excitation of carriers in the QWR, while the plateau  can be explained by the diffusion and relaxation of barrier-QW carriers into the QWR states. A similar process was reported for carrier transfer in the nanosecond timescale between QWs and  sidewall QWRs on GaAs(113)A substrates\cite{Lienau98}. Note that although the bulk emission is spectrally closely located to the QWR, its fast decay implies only a minor contribution to the PL signal.


\subsection{Acoustic Transport}
\label{sec:transport}

\begin{figure*}[hbtp]
\includegraphics[width=.9\textwidth, angle=0, clip]{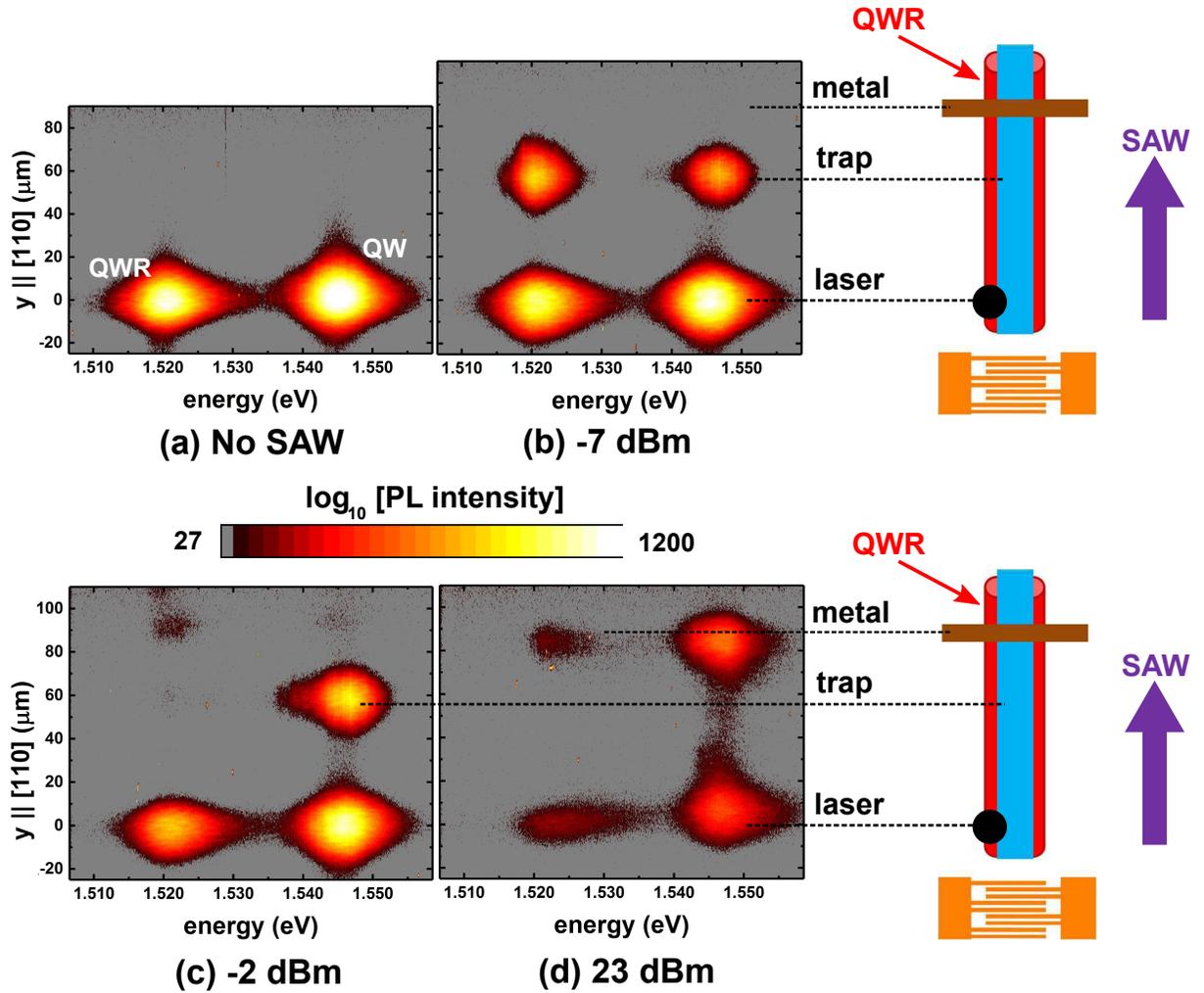}
\caption{Acoustic transport of photoexcited charge carriers in sidewall QWRs and surrounding QW. Photoluminescence maps recorded using the configuration in the right panels without SAW (a) and with SAWs with nominal powers of (b) -7 dBm, (c) -2 dBm, and (d) 23 dBm. The carriers are excited with a 635 nm  laser focused to a spot of approximately \micron{6} diameter on a ridge sidewall. The carriers are transported by the SAW field to a trapping center or, for high acoustic powers, up to a metal stripe, which induces recombination by quenching the SAW piezoelectric field.}
\label{fig:Fig6}
\end{figure*}

We provide in this section experimental evidence for acoustic charge transport along the sidewall QWRs, obtained by spatially resolved PL spectroscopy. The experiments were carried out  using the configuration sketched in the right panel of Fig.~\ref{fig:Fig6}. The carriers were photo-excited by a pulsed diode laser (\nanometer{635}, power \SI{2}{\micro\watt}) focused on to a spot with a diameter of \micron{6} at the reference position $x= \micron{0}$.  A metal stripe was  deposited at the end of the QWR (at $y=\micron{90}$) to block the acoustic transport. Since the depth of the QWR (below a \nanometer{200} AlGaAs barrier) is much smaller than the SAW wavelength ($\lSAW=5.6~\mu$m), the metal efficiently screens the SAW piezoelectric potential at the depth of the QWR and induces the recombination of the carriers transported by the SAW field \cite{PVS135}. In the ideal case, one thus expects to observe PL at two positions along the SAW path: (i) at the excitation location, where non-transported carriers recombine and (ii) at the metal stripe position due to recombination of transported carriers. However, as will be shown in the following, we observe PL from positions along the SAW path, which are attributed to unintended trapping centers. The centers capture carriers of one polarity, which then recombine upon the arrival of carriers of opposite polarity in the subsequent SAW half-cycle. Furthermore, in the present case,  the metal stripe consists of a stack of \nanometer{10} titanium, \nanometer{30} aluminium and \nanometer{10} titanium layers. Using the optical transmission data taken from reference \cite{RefrIndex}, we calculate an optical transmittance of only \SI{0.8}{\percent} at the QWR emission ($\lambda=$\nanometer{815.8}). We expect, therefore, to collect only the PL emitted by carriers recombining along the transport path and at the edges of the metal stripe. The small optical transmittance of the metal stripe should suppress PL from directly below the metal stripe.
 
In the absence of SAW excitation, PL  is only observed around the excitation spot (Fig.~\ref{fig:Fig6}(a)). The application of a SAW of \SI{-7}{\decibel\metre} induces electron-hole transport in the QW and the QWR, as illustrated in Fig.~\ref{fig:Fig6}(b). This image shows a strong remote PL emission  at a position of \micron{60} away from the excitation spot, which we attribute to carrier trapping and recombination in a center within the transport channel. Interestingly, the remote PL is observed in the same position for both the QWR and the QW resonances. We remind ourselves that carrier transfer between the QW and the QWR is hindered by the barrier-QW in-between them. In addition, we have shown in Sec.~\ref{sec:optical} that only carriers generated in the barrier QW can diffuse to the QWR, and that this process takes place within a nanosecond. Considering the acoustic transport velocity of \SI{2.9}{\micro\metre\per\nano\second}, this transfer between the barrier QW and the QWR can only take place during the first couple of micrometers of the transport path. The simultaneous remote emission from the QW and QWR could be due to an extended defect affecting both the QW and the QWR. Alternatively, if the acoustic transport in the QWR is blocked by a defect center,  carriers can  accumulate close to the defect and eventually leak to the QW. 

A further interesting aspect is that trapping and recombination along the transport channel can be controlled by the acoustic intensity. The images in Figs.~\ref{fig:Fig6}(b) and  \ref{fig:Fig6}(c) show that an increase of the acoustic power to \SI{-2}{\decibel\metre} leads initially to an enhancement of the trap emission. After a certain power level, however, the emission from the trap becomes completely suppressed (Fig.~\ref{fig:Fig6}(d)). The SAW piezoelectric field becomes, in this case, sufficiently strong  to overcome the trapping potential and transport the carriers all the way to the metal stripe 90~$\mu$m away from the excitation spot. Note that the SAW power threshold for overcoming the trap potential is different for transport in the QWR and QW. At the metal stripe position, the piezoelectric field is screened, leading to the recombination of transported carriers. However, as was mentioned before, due to the small optical transmission of \SI{0.8}{\percent} of the metal stripe, we expect to only observe a small fraction of the PL at this position. Additionally, for very high SAW powers the piezoelectric field is only partially screened and therefore the carriers are potentially transported further along the SAW path (out of the measurement area). 

\begin{figure}[htbp]
\includegraphics[width=1\columnwidth, angle=0, clip]{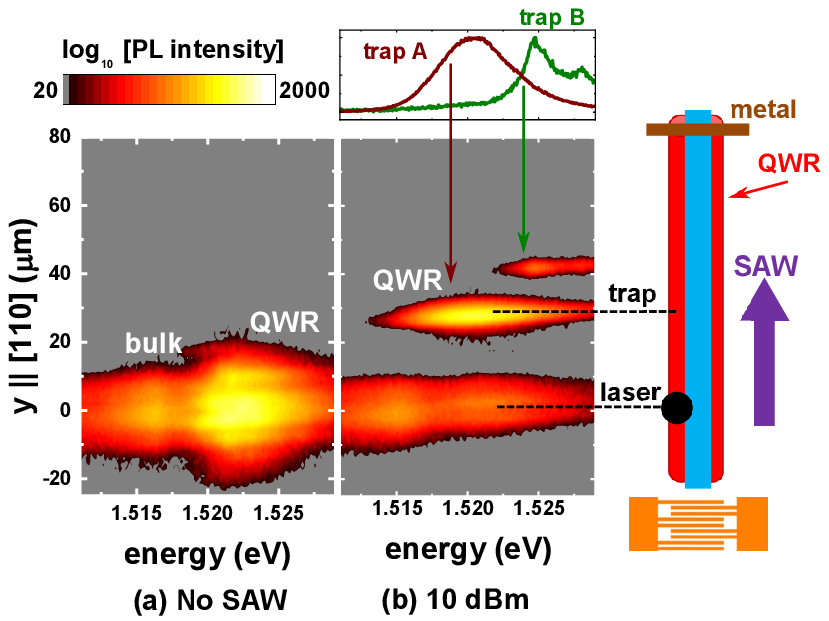}
\caption{Spatially resolved PL maps recorded (a) in the absence of a SAW and (b) under a SAW with wavelength of  \SI{726}{\mega\hertz} and  nominal rf excitation power of 10~dBm. The carriers were selectively excited  in the QWR  at $y=0$ (cf. right panel) using a laser beam tuned to an energy ($E_\mathrm{laser}=$\ev{1.537}) below the band gap of the surrounding QW. Under acoustic excitation, the carriers are transported along the wire and recombine in two trapping centers (A and B) with spectral emissions displayed in the upper panel.
}
\label{fig:Fig7}
\end{figure}

Figure~\ref{fig:Fig7} displays PL maps recorded by selectively exciting carriers in the QWR using an excitation laser energy (\ev{1.537}) below the QW resonance. In this case, the acoustic transport takes place only in the QWR. In the absence of a SAW (Fig.~\ref{fig:Fig7}(a)) PL from the QWR is only observed at the excitation spot near a weak emission line from the GaAs substrate. When a SAW is applied ($\lambda_{SAW} = $\micron{4}, $f_{SAW} = $\SI{726}{\mega\hertz}, cf. Fig.~\ref{fig:Fig7}(b)), one detects PL at two remote locations on the QWR, thus evidencing acoustic transport. The upper panel displays PL spectra of the two trapping sites. The spectrum from trap A (red) resembles the  QWR emission at the excitation spot (\ev{1.521}). A similar behaviour has been found for most of the trapping centers - a model of this type of trap centers will be presented below. In contrast, the trapping center at site B shows a narrower emission line (linewidth of \mev{2} in comparison with \mev{7} for trap A), which is blue-shifted by \mev{4} with respect to the PL at the excitation location. This indicates that trap B is caused by a different type of trap. 

\begin{figure}
\includegraphics[width=1\columnwidth, angle=0, clip]{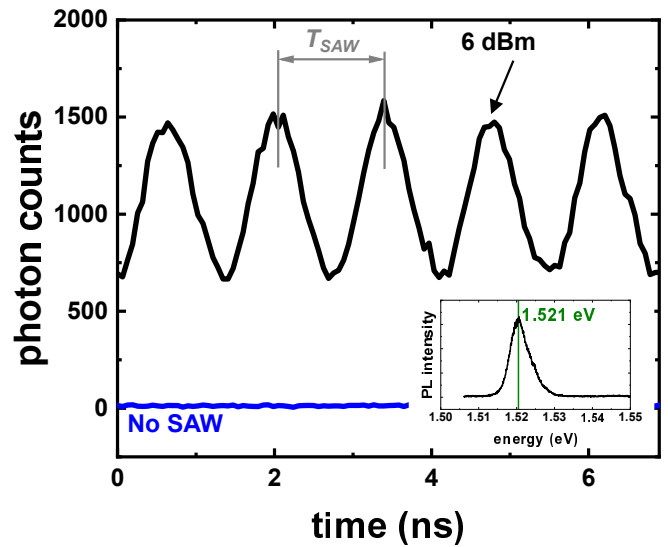}
\caption{Time-resolved remote PL  measured on a trap with the spectral emission displayed in the inset. The trap was populated by carriers transported by a SAW with a wavelength of \micron{4}, frequency \SI{726}{\mega\hertz} and nominal rf excitation power of \SI{6}{\decibel\metre}. The blue curve shows, for comparison, the PL trace in the absence of a SAW.
}
\label{fig:Fig8}
\end{figure}

The dynamics of the acoustic transport was studied by time-resolved PL. The carriers were optically excited using a small spot (diameter of  approximately \micron{3}) using a cw Ti:Sapphire laser (wavelength of \nanometer{780} and power of \SI{200}{\micro\watt}).  The remote PL induced by the acoustic transport of carriers to a  trap within the QWR was collected and detected using a Si avalanche detector (temporal resolution of \SI{340}{\pico\second}) synchronized with the SAW phase.
Figure~\ref{fig:Fig8} shows the PL time dependence recorded on a trap located approximately \micron{30} away from the excitation spot. The cw PL spectrum of the trap  (inset) reveals that it is similar to trap A in Fig.~\ref{fig:Fig7}(b) emitting at the same energy as the  QWR. The PL intensity oscillates at the SAW frequency of  \SI{726}{\mega\hertz}.  For comparison, the blue curve displays the vanishing PL intensity collected in the absence of SAW, which proves that the trap is only populated by acoustically transported carriers.

The strong oscillations in the  time-resolved trace of Fig.~\ref{fig:Fig8} proves that the carriers remain confined within the SAW potential during acoustic transport (note that the amplitude of the oscillations in Fig.~\ref{fig:Fig8} is partially limited by the time resolution of the avalanche detector of approximately \SI{340}{\pico\second}). In addition, the frequency of the oscillations yields information about the carrier trapping and recombination process. In fact, one expects PL oscillations at the SAW frequency only if the trapping center captures carriers of only one polarity during one half of the SAW cycle, followed by recombination upon arrival of carriers with the opposite polarity in the subsequent half of the SAW cycle. The capture (and subsequent recombination) of carriers of both polarities results, in contrast,  in  PL oscillations at twice the SAW frequency. In the present experiments, we cannot discriminate which carriers (electrons or holes) are preferentially trapped. 

Our experimental data supports the model for carrier trapping and recombination during ambipolar transport proposed in Ref.~\onlinecite{PVS244}. In this model, carriers of one polarity are trapped due to injection into states at the GaAs/AlGaAs interface  by the vertical component $E_z$ of the SAW piezoelectric field. The field-induced injection probability is high since, for a Rayleigh SAW,  $E_z$ is in phase with the piezoelectric potential and, therefore, with the maximum in the carrier density. $E _z$ reverses its sign half a SAW-cycle later, thus releasing the trapped carriers into a pool of carriers of the opposite polarity, where it has a high recombination probability. This mechanism thus leads to efficient recombination during acoustic transport with emission energy equal to the QWR. This model can account for the emission behaviour of most of the trapping sites detected along the acoustic transport path (i.e., of type A, cf. Fig.~\ref{fig:Fig7}).

\section{\label{sec:conclusion}Conclusions}
We have studied structural, optical and transport properties of sidewall QWRs fabricated on patterned GaAs(001) substrates. QWRs formed by overgrowth of a \nanometer{10}-wide QW on pre-patterned ridges have thickness and width of approximately \nanometer{25} and \nanometer{200}. The sidewall QWRs form on both sides of the ridges aligned along the [110] crystal direction and exhibit similar emission properties at both sides of the ridges. Their emission is spectrally red-shifted by \mev{27} with respect to the fundamental transition of the QW. The linewidth of the QWR emission lines of \mev{10.5} is comparable to the one observed for the QW. One source of potential fluctuations contributing to the QWR linewidth are fluctuations in the lateral dimensions of the QWRs, as is implied by the increased linewidth of the QWR with respect to the QW. By combining S(T)EM and AFM measurements, we estimate that the amplitude of line-edge-roughness is small as compared to the QWR lateral width. LER is therefore expected to play only a minor role in the acoustic transport of carriers. Furthermore, LER is mainly caused by the photolithography process.

We have further demonstrated that photo-excited electrons and holes can be acoustically transported along the QWR over distances of at least \micron{90}. By selectively exciting carriers using a laser energy below the band gap of the surrounding QW, we proved that the majority of the QWR carriers is transported along the QWR and not via the surrounding QW. The transport distances are in many cases limited by the presence of trapping centers along the transport channel, which block the transport and induce recombination of the carriers.  These traps have been observed both in the QW and the sidewall QWR. The majority of the traps in the QWR exhibits emission energies and linewidths comparable to the ones expected for the QWR. Some of the traps can, however, exhibit much narrower linewidth (down to approx. \mev{2}) and a slightly different emission energy. The submicrometer widths, the efficient lateral confinement and acoustic transport properties of the sidewall QWRs on GaAs(001) substrates make them a promising candidate for interconnections of charge and spin carriers in semiconductor structures.

\section{Acknowledgements}
We thank Jens Herfort for a critical reading of the manuscript, as well as W. Anders, A. Tahraoui, and S. Rauwerdink  for the help in the fabrication of the samples. This project has received funding from the European Union's Horizon 2020 research and innovation program under grant agreement No 642688.

\bibliography{literature,mypapers}

\end{document}